\documentstyle[epsfig,aps,twocolumn]{revtex}

\begin{document}
\title{Optical realization of universal quantum cloning}
\author{Yun-Feng Huang, Wan-Li Li, Chuan-Feng Li\thanks{%
Electronic Address: cfli@ustc.edu.cn}, Yong-Sheng Zhang, Yun-Kun Jiang,
Guang-Can Guo\thanks{%
Electronic Address: gcguo@ustc.edu.cn}}
\address{Laboratory of Quantum Communication and Quantum Computation and Department\\
of Physics, University of Science and Technology of China,\\
Hefei, 230026, P. R. China\vspace{0.3in}}
\maketitle

\begin{abstract}
Beyond the no-cloning theorem, the universal symmetric quantum cloning
machine was first addressed by Bu\v {z}ek and Hillery. Here, we realized the
one-to-two qubits Bu\v {z}ek-Hillery cloning machine with linear optical
devices. This method relies on the representation of several qubits by a
single photon. We showed that, the fidelities between the two output qubits
and the original qubit are both $\frac 56$ (which proved to be the optimal
fidelity of one-to-two qubits universal cloner) for arbitrary input pure
states.

PACS number(s): 03.67.-a, 03.65.Bz
\end{abstract}

\section{Introduction}

A deterministic and perfect quantum cloner for arbitrary quantum states is
forbidden by the quantum no-cloning theorem\cite{h01}. As an extension, Yuen
and D'Ariano\cite{h02} showed that a violation of unitarity makes the
cloning of two nonorthogonal states also impossible. Then the case of mixed
states is considered and it is shown that two noncommuting mixed states
cannot be broadcast, even when the state need only be reproduced marginally
on the separate systems. For further consideration, Koashi and Imoto\cite
{h03} generalized the standard no-cloning theorem to the entangled states.
Their research showed to what extent the correlation to other systems can be
read out from a subsystem without altering its marginal density operators.

After having considered about so many restrictions posed on cloning quantum
states by quantum mechanics, people began to think about how to do our best
in quantum states cloning under all these restrictions, that is, the
inaccurate cloning of quantum states. These approached may be divided into
two main categories: universal and state-dependent.

State-dependent quantum cloning machines are designed to clone states
belonging to a finite set. It can be divided into three subcategories:
deterministic\cite{h04}, probabilistic exact\cite{h05} and hybrid\cite{h06}.

On the other hand, universal quantum cloning machines deterministically
generates approximate copies of an unknown quantum state. The word
``universal'' means that this kind of machines works equally well for any
input states, i.e., state-independent. This category can again be divided
into two subcategories: symmetric and asymmetric.

Asymmetric universal quantum cloning machines which introduced by Cerf and Bu%
\v {z}ek et al\cite{h07,h08} produce two non-identical imperfect copies of a
single state of $N$-dimensional quantum system. Symmetric universal quantum
cloning machines, first designed by Bu\v {z}ek and Hillery\cite{h04}, works
in a different way: act on any unknown quantum state and produce identical
approximate copies equally well. This kind of Bu\v {z}ek-Hillery cloning
machine has been optimized and generalized\cite{h09,h10}.

Quantum cloning may be used to improve the performance of some quantum
computation tasks\cite{h19} and may be heleful in understanding quantum
information theory and processing quantum imformation, and might be helpful
in eavesdropping versus quantum cryptography\cite{h20}. And we can improve
the performance of the measurements on observables by performing
measurements on clones of the original quantum system\cite{h21}.

In the present paper, we realized the one-to-two qubits Bu\v {z}ek-Hillery
cloning machine with linear optical devices. This method relies on the
representation of several qubits by a single photon. We showed that, the
fidelities between the two output qubits and the original qubit are both $%
\frac 56$ (which proved to be the optimal fidelity of one-to-two qubits
universal cloner) for arbitrary input pure states.

\section{Universal cloning}

The most important quality of symmetric universal cloning machines is how
well we can do, i.e., the optimal fidelity. We use the most usual definition
of fidelity $F=\left\langle \Psi \left| \rho \right| \Psi \right\rangle $.
For convenience, we only discuss the case when the input states are pure.
Here,$\,|\Psi \rangle $ is the to-be-cloned unknown pure state and $\rho $
is the density matrix of one of the output copies.

A $M\rightarrow N$ symmetric universal quantum cloning machine is described
as the following (for qubit systems):

1. The input system contains $M$ original qubits, each in the same unknown
pure state $|\Psi \rangle $, $N-M$ ``blank'' qubits $|0\rangle $ and $K$
auxiliary qubits $|0\rangle $.

2. The cloning machine acts on the input system and performs a prescribed
unitary transformation $U$ on it.

3. The output system contains $N$ identical copies (usually entangled), each
copy described by a reduced density matrix $\rho _i^{out}$ $(i=1,\cdot \cdot
\cdot ,N)$, and $K$ auxiliary qubits.

The cloning process can be written as

\begin{equation}
U|\Psi \rangle ^{\otimes M}|0\rangle ^{\otimes (N-M)}|aux\rangle =|\Phi
\rangle ^{out}\text{.}  \eqnum{1}
\end{equation}
For $|\Phi \rangle ^{out}$, there exists $N$ qubits such that each of them
has the same reduced density matrix $\rho ^{out}$.

For the $M\rightarrow N$ symmetric universal quantum cloning machine
described above, the optimal fidelity has been demonstrated as\cite{h10}

\begin{equation}
F^{opt}(M,~N)=\frac{MN+N+M}{N(M+2)}\text{.}  \eqnum{2}
\end{equation}
From the equation we can see that the optimal $1\rightarrow 2$ fidelity is $%
\frac 56$.

The special unitary transformation $U$ of the one-to-two Bu\v {z}ek-Hillery
cloning machine can be written as\cite{h04} 
\begin{eqnarray}
&&|0\rangle _1|0\rangle _2|0\rangle _3\stackrel{U}{\longrightarrow }\sqrt{%
\frac 23}|00\rangle _{12}|0\rangle _3+\sqrt{\frac 13}|+\rangle
_{12}|1\rangle _3\text{,}  \eqnum{3} \\
&&|1\rangle _1|0\rangle _2|0\rangle _3\stackrel{U}{\longrightarrow }\sqrt{%
\frac 23}|11\rangle _{12}|1\rangle _3+\sqrt{\frac 13}|+\rangle
_{12}|0\rangle _3\text{,}  \nonumber
\end{eqnarray}

where

\begin{equation}
|+\rangle =\sqrt{\frac 12}(|10\rangle _{12}+|01\rangle _{12})  \eqnum{4}
\end{equation}

In the cloning process, the quantum information in the original qubit (the
first qubit) is copied to qubits $1$ and $2$ equally well such that the
reduced density matrixes at the output are identical, i.e., $\rho
_1^{out}=\rho _2^{out}$. The third qubit acts as an auxiliary qubit here.

From the specific expression of the unitary transformation $U$ given above
we can easily calculate the fidelities 
\begin{eqnarray}
F_1 &=&\langle \Psi |\rho _1^{out}|\Psi \rangle =\frac 56\text{,}  \eqnum{5}
\\
F_2 &=&\langle \Psi |\rho _2^{out}|\Psi \rangle =\frac 56\text{,}  \nonumber
\end{eqnarray}

where $|\Psi \rangle $ is an arbitrary unknown pure state

\begin{equation}
|\Psi \rangle =\alpha |0\rangle +\beta e^{i\varphi }|1\rangle ,\text{ }%
\alpha ,\beta \in real,\text{ }\varphi \in [0,2\pi ).  \eqnum{6}
\end{equation}
So, it is shown to be an optimal symmetric universal cloning machine.

In order to realize the transformation $U$ in experiment, we first have to
construct the quantum logic circuit. To do that, $U$ is decomposed into a
sequence of basic operations such as the rotation of a single qubit and the
controlled-NOT (CNOT) operation of two qubits\cite{h22}. The final network
of the cloning machine\cite{h11} has been shown in figure 1.

We note that the network is divided into two parts (except the state
swapping part): preparation and cloning. In the preparation stage, qubit 1
is left unaffected and qubit 2 and 3 are entangled by CNOT gates. That is,
there is no ``flow'' of quantum information from qubit 1 to qubit 2, 3.
Then, in the cloning stage, the quantum information in qubit 1 is
``redistributed'' among the three qubits by a sequence of four CNOT gates.

The unitary transformations in the two stages are described as\cite{h11}

\begin{equation}
|\Psi \rangle _1|0\rangle _2|0\rangle _3\stackrel{U_1}{\longrightarrow }%
|\Psi \rangle |\Psi \rangle _{23}^{prep}\stackrel{U_2}{\longrightarrow }%
|\Psi \rangle _{123}^{out}\text{,}  \eqnum{7}
\end{equation}
where $|\Psi \rangle _{23}^{prep}$is prepared by three ratations and two
CNOT operations

\begin{equation}
|\Psi \rangle _{23}^{prep}=\hat{R}_2(\theta _3)\hat{P}_{32}\hat{R}_3(\theta
_2)\hat{P}_{23}\hat{R}_2(\theta _1)|0\rangle _2|0\rangle _3.  \eqnum{8}
\end{equation}
The rotation angles $\theta _j$ $(j=1,2,3)$ are determined by the specific $%
|\Psi \rangle _{23}^{prep}$. In this case the state $|\Psi \rangle
_{23}^{prep}$ reads

\begin{equation}
|\Psi \rangle _{23}^{prep}=\sqrt{\frac 16}(2|00\rangle _{23}+|01\rangle
_{23}+|11\rangle _{23}).  \eqnum{9}
\end{equation}

Although there is no flow of information in the preparation stage, we note
that we can controll the distribution of information among the three qubits
by the choice of the ratation angles\cite{h11}. We can copy qubit 1 to
qubits 1 and 2 or qubits 2 and 3, even copy qubit 1 to qubits 1, 2 and 3
when the input state $|\Psi \rangle $ is

\begin{equation}
|\Psi \rangle =\alpha |0\rangle +\beta |1\rangle ,\text{ }\alpha ,\beta \in
real.  \eqnum{10}
\end{equation}
This is called quantum triplicator.

\section{Experiment description}

Now we concentrate on constructing the network given in figure 1 with linear
optical devices. The basic idea is the representation of several qubits by a
single photon. In such a scheme, qubits are represented by different freedom
degrees of a single photon, i.e., a photon can carry one polarization qubit
and several location qubits\cite{h12,h13,h14}. Two orthogonal linear
polarization states of the photon, i.e., horizontal and vertical, serve as
the basis states of the polarization qubit. Hereafter, they are denoted by $%
\left| H\right\rangle $ and $\left| V\right\rangle $ corresponding to $%
\left| 0\right\rangle $ and $\left| 1\right\rangle $, respectively. Any of
the location qubits is characterized by the information about ``which path''
is taken by the photon when passing a beam splitter. In this way,
entanglement of different qubits can be accomplished conveniently, and
decoherence is very low because the photons have relatively less interaction
with the environment.

In our scheme of universal quantum cloning, three qubits are involved: one
polarization (qubit $1$) and two location qubits (qubits $2$ and $3$). Qubit 
$1$ is the initial to-be-cloned qubit in an arbitrary state $\left| \Psi
\right\rangle _1$, which is cloned to qubits $1$ and $2$ at the output end
with qubit $3$ serving as an ancilla. Both location qubits are initially in
blank states: $\left| 0\right\rangle _2$ and $\left| 0\right\rangle _3$.

To realize the network in figure 1, we must construct CNOT gates between any
two qubits and unitary rotation of any single qubit. With the representation
of qubits mentioned above, all these gates can be made with linear devices%
\cite{h13,h15}. However, in previous schemes\cite{h15}, a unitary rotation
to a single location qubit is realized by a shiftable beam splitter (BS) or
a Mach-Zehnder (M-Z) interferometer. Unfortunately, a shiftable BS is
infeasible and an M-Z interferometer will inevitably multiply the complexity.

To overcome this difficulty, a trick of state-swapping between qubits $1$
and $2$ is employed to transform the input state $\left| \Psi \right\rangle
_1\left| 0\right\rangle _2\left| 0\right\rangle _3$ to $\left|
H\right\rangle _1\left| \Psi \right\rangle _2\left| 0\right\rangle _3$. Then
with a definite polarization state, unitary rotations to a single location
qubit can be realized by an operation of Controlled-Rotation (polarization
controls location), which can be implemented by only using one Polarization
Beam Splitter (PBS) and several properly oriented Half-Wavelength Plates
(HWP).

After the state-swapping, we can begin to construct the two parts of our
network: preparation and cloning. After having finished this work, we may
have a complicated construction (composed of many universal logic gates) of
such a cloning machine with optical deveces. But by techniques of compiling%
\cite{h14}, the network can be significantly simplified and a feasible setup
is shown in figure 1.
\begin{figure}
\epsfig{file=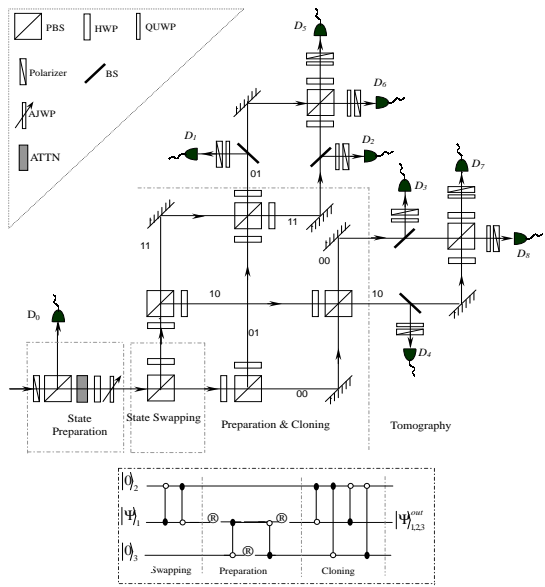,height=10.5cm,width=7.8cm}
\caption{the setup of the universal quantum cloning machine.}
\end{figure}

Now let's explain the experiment setup (figure 1) in detail.

The performance of this cloning machine is demonstrated by using a laser
beam attenuated to the single-photon level as input source. At the output,
the photons are detected using geigermode avalanche photodiodes --- Single
Photon Counting Modules ($D1$-$D8$, EG\&G \#SPCM-AQ, efficiency $\sim 70\%$,
dark count $\sim 50s^{-1}$). In our experiment, a well polarized light beam
is created by a He-Ne laser with a Brewster window (Melles Griot, $633$ $nm$
) and its polarization degree is improved by a polarizer. Then a PBS
follows, and its vertical output is measured by a power meter $D_0$ to
confirm the stability of the laser's power during the experiment. The
horizontal output of this PBS is greatly attenuated (by an attenuating
plate, ATTN) so that the maximum detection rates are always less than $%
20000s^{-1}$; for the setup passage time of $5$ $ns$, this means that on
average fewer than $10^{-4}$ photons are in the setup at any time. Now the
input source can be viewed as single-photon source\cite{h16}.

After that the to-be-cloned state is prepared by a properly oriented HWP and
an adjustable waveplate (AJWP, namely, a Pockels Cell).

To confirm the correct result, in a general way, quantum state tomography%
\cite{h16} at the output end is required, and a simple version by optical
interference is considered in the present system. Another auxiliary qubit is
utilized to determine the reduced density operators of the two replicas
respectively. This auxiliary qubit (qubit {\it aux}) is introduced as $\frac %
1{\sqrt{2}}(|0\rangle _{aux}+|1\rangle _{aux})$ by four $50-50$ BSes. Then a
controlled state swapping is performed with qubit {\it aux }as the
controller: if qubit {\it aux is }$|1\rangle _{aux}$, the states of qubits $%
1 $ and $2$ interchange, or else they remain unchanged. The four paths with
qubit {\it aux }as $|0\rangle _{aux}$ determine the density operator of
qubit $1$ (polarization): if the relative photon counts and the polarization
state of path $i$ is $C_i$ and $\rho _i$ respectively (by measuring four
linearly independent states of the polarization qubit: $|H\rangle ,$ $%
|V\rangle ,$ $|D\rangle ,$ $|R\rangle $, we can determine each component of
the matrix $\rho _i$, here $|D\rangle =\frac 1{\sqrt{2}}(|0\rangle
+|1\rangle ),$ $|R\rangle =\frac 1{\sqrt{2}}(|0\rangle +e^{i\frac \pi 2%
}|1\rangle )$,and this method is called quantum tomography\cite{h16}), the
density operator of qubit $1$ can be written as $\sum C_i\rho _i$. Other
four paths determine the density operator of qubit $2$ (location) which has
been swapped to the polarization qubit with {\it aux }as $|1\rangle _{auz}$,
and the method of measurement is the same as above.

Then the reduced density operators of the replicas provide the evaluation of
the fidelities,which are shown in Figure 2. We can see that the experiment
datas coincide with the theoretical value $\frac 56$ quite well.

Before qubit $2$ is under verification, it has been swapped with the
polarization one, so errors of both qubits' fidelities can be analyzed in
similar ways. Now we will analyze the error of our experiment. The error of
the fidelity $F$ originates mainly from two ways:

1. The oscillation of the photon counts of pass $i$ (denoted by $\Delta C_i$%
), which is primarily caused by the small oscillation of the reflexivity and
transmissivity of the first HWP (in the stage of state preparation) for
different polarization states.

2. The precision limits of the waveplates and polarizers' orientation angles
(denoted by $\Delta \theta $).

It can be deduced that an upper bound of the maximum error of $F$ can be
expressed as\cite{h17} $\Delta F=\sum\limits_{i=1}^4\Delta C_i+\frac 32%
\Delta \theta $. In the present system, $\Delta \theta \approx 0.0018$ $rad$
($0.1^{\circ }$) and $\sum\limits_{i=1}^4\Delta C_i\approx 0.002$. Thus we
get the maximum error of $F$ as $\Delta F\approx 0.005$, which has been
embodied in the result figure as error bars.
\begin{figure}
\epsfig{file=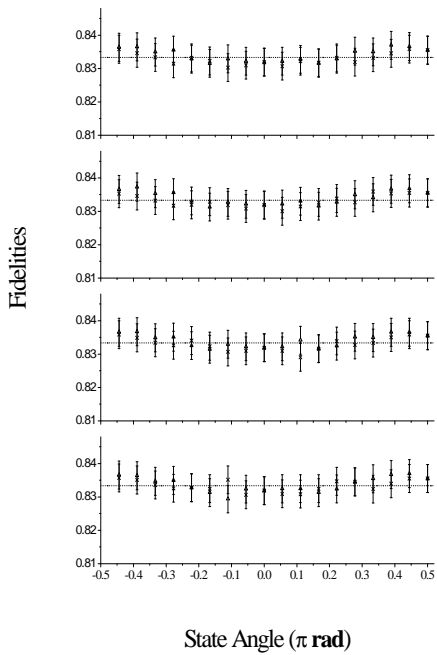,height=12cm,width=8cm}
\caption{the replicas' fidelities in the optimal universal cloning machine.}
\end{figure}

A more complicated network can be designed to carry out the $N\rightarrow M$
cloning, whose principle has been generalized from the $1\rightarrow 2$
cloning\cite{h04,h10}. But in the present scheme, qubits are represented by
different freedom degrees of a single photon, and about $2^n$ optical
elements are needed to represent $n$ qubits\cite{h13}. So the complexity of
the network will increase exponentially with the increase of $N$ and $M$,
and this is a shortcoming of the present method.

In the present experiment, qubits are represented by different freedom
degrees of a single photon. With a quantum non-demolition (QND) measurement%
\cite{h18}, the information of the location qubit may be extracted without
destroying the polarization qubit, in principle.

\section{Setup and Results}

Figure 1 is the setup of the universal quantum cloning machine. The inset is
the logic circuit ($\bullet $ and $\circ $ denote the controller and target
of a CNOT operation respectively, $R$ is a rotation to a single qubit). We
represent the three input qubits by three freedom degrees of a single
photon. Qubit $1$ (to-be-cloned polarization qubit) is introduced by setting
the photons' initial polarization in an arbitrary state $\left| \Psi \left(
\theta ,\delta \right) \right\rangle _1=\cos \theta \left| H\right\rangle
_1+e^{i\delta }\sin \theta \left| V\right\rangle _1$ $\left( -\frac \pi 2%
<\theta \leq \frac \pi 2,\text{ }0\leq \delta <\pi \right) $. Qubits $2$ and 
$3$ are both location qubits, which are introduced by different choices of
``beam paths'' with the initial state $\left| 00\right\rangle _{2,3}$. By
the state-swapping, qubit $2$ is transformed to $\left| \Psi \left( \theta
,\delta \right) \right\rangle _2=\cos \theta \left| 0\right\rangle
_2+e^{i\delta }\sin \theta \left| 1\right\rangle _2$, while qubit $1$ is
transformed to $\left| H\right\rangle _1$. In the stage of entanglement
preparation, qubits $1$ and $3$ are prepared into $\left| \Psi \right\rangle
_{1,3}^{prep}=\frac 1{\sqrt{6}}\left( 2\left| H0\right\rangle _{1,3}+\left|
H1\right\rangle _{1,3}+\left| V1\right\rangle _{1,3}\right) $. Then several
CNOT operations follow to accomplish the cloning, which can be compiled to
two independent polarization M-Z interferometers. Quantum state tomography
is adopted to confirm the correct result. Another auxiliary qubit enables us
to measure the density operators of the replicas separately, which is
introduced as $\frac 1{\sqrt{2}}\left( \left| 0\right\rangle _{aux}+\left|
1\right\rangle _{aux}\right) $ by arranging four $50${\it -}$50$ BSes in
each potential paths of the output photon. Then a controlled state swapping
is carried out and the four qubits (qubit $1,$ $2,$ $3$ and $aux$) evolve to
the state $\left| \Psi \right\rangle ^{meas}=$ $\frac 1{\sqrt{2}}\left(
\left| 0\right\rangle _{aux}\left| \Psi \right\rangle _{1,2,3}^{out}+\left|
1\right\rangle _{aux}\left| \Psi \right\rangle _{2,1,3}^{out}\right) $. When
the qubit $aux$ is $\left| 0\right\rangle _{aux}$, the density operator of
replica qubit $1${\it \ }(polarization) can be obtained by detectors $1$ to $%
4$ ($D_1${\it -}$D_4$) and corresponding QUWPs and polarizers, by measuring
the mixed state of polarization and relative photon counts of each path. To
the other four outputs of the BSes where the qubit $aux$ is $\left|
1\right\rangle _{aux}$, a state-swapping between qubits $1$ and $2$ is
performed so that the density operator of replica qubit $2$ is exactly
realized by knowing that of the interchanged polarization qubit, which is
measured by $D_5${\it -}$D_8$ and corresponding QUWPs and polarizers.

Figure 2 is the replicas' fidelities in the optimal universal cloning
machine. The fidelity is defined as $\left\langle \Psi \left( \theta ,\delta
\right) \right| \rho \left| \Psi \left( \theta ,\delta \right) \right\rangle 
$ with $\rho $ denoting the density operator of a replica. This machine can
deal with any state $\left| \Psi \left( \theta ,\delta \right) \right\rangle
=\cos \theta \left| V\right\rangle +e^{i\delta }\sin \theta \left|
H\right\rangle $ equally well. From down to up, the four subfigures
represent the cases $\delta =0,\frac \pi 4,\frac \pi 2,\frac{3\pi }4$,
respectively. In each subfigure, the state angle $\theta $ of the
to-be-cloned qubit varies from $-\frac \pi 2$ to $\frac \pi 2$, the dotted
lines represent the theoretical optimal fidelities $\frac 56$ $\left(
\approx 0.833\right) $, while the crosses $\left( \times \right) $ denote
the fidelities of the output qubit $1$ and the triangles $\left(
\bigtriangleup \right) $ denote the fidelities of qubit $2$. As is shown,
all the fidelities are close to the theoretical value $\frac 56$, and errors
have been analyzed in section III.

\section{Conclusion}

We have realized a one-to-two qubits Bu\v {z}ek-Hillery cloning machine with
linear optical devices. We showed that, the fidelities between the two
output qubits and the original qubit are both $\frac 56$ for arbitrary input
pure states.

{\bf Acknowledgment} This work was supported by the National Natural Science
Foundation of China.

\begin{references}
\bibitem{h17}  Considering the output state of the polarization qubit as $%
\rho =\sum\limits_{i=1}^4C_i\left| \Psi \left( \theta _i,\delta _i\right)
\right\rangle \left\langle \Psi \left( \theta _i,\delta _i\right) \right| $,
the fidelity is calculated by $F=\left\langle \Psi \left( \theta ,\delta
\right) \right| \rho \left| \Psi \left( \theta ,\delta \right) \right\rangle 
$. Let $F=\sum\limits_{i=1}^4f_i$, where $f_i=\frac 12C_i\left[ 1+\cos
2\theta \cos 2\theta _i+\sin 2\theta \sin 2\theta _i\cos \left( \delta
_i-\delta \right) \right] $. It is easy to prove that $\Delta f_i\leq \Delta
C_i+C_i\left( \Delta \theta _i+\frac 12\Delta \delta _i\right) $, and the
error of $F$ can be expressed as $\Delta F=\sum\limits_{i=1}^4\Delta f_i$.
Then the maximum error may be written as $\Delta F=\sum\limits_{i=1}^4\Delta
C_i+\frac 32\Delta \theta $, for $\sum\limits_{i=1}^4C_i=1$ and $\Delta
\theta _i=\Delta \delta _i=\Delta \theta $ is the orientation precision.
\end{references}
\end{document}